\documentclass[a4paper, 11pt]{article}

\usepackage{amsmath,pifont,amssymb}
\usepackage{psfrag}
\usepackage{subfigure}
\usepackage{url}
\usepackage{color}
\usepackage[latin1]{inputenc}
\usepackage[T1]{fontenc}
\usepackage[english]{babel}
\usepackage{verbatim}
\usepackage{graphics,array,float,epsfig}
\usepackage{amsfonts,graphicx,color}
\usepackage{dsfont}
\usepackage{mathrsfs}
\usepackage{stmaryrd}
\usepackage{authblk}

\newcommand{\beq}{\begin{equation}}
\newcommand{\eeq}{\end{equation}}
\newcommand{\beqs}{\begin{subequations}}
\newcommand{\eeqs}{\end{subequations}}
\newcommand{\benum}{\begin{enumerate}[label=(\roman*)]}
\newcommand{\eenum}{\end{enumerate}}

\newcommand{\W}{\boldsymbol{W}}
\newcommand{\dif}{{\mathrm{d}}}

\newcommand{\defin}{\stackrel{\scriptscriptstyle\triangle}{=}}

\newcommand{\w}{\boldsymbol{u}}
\newcommand{\ws}{\boldsymbol{{u}_{a}}}

\newcommand{\bu}{\boldsymbol{u}}
\newcommand{\bus}{\boldsymbol{{u}_{a}}}
\newcommand{\buss}{\boldsymbol{{u}^\circ_{a}}}
\newcommand{\bv}{\boldsymbol{v}}

\newcommand{\bsigma}{\boldsymbol{\sigma}}

\newcommand{\XX}{\mathbf{x}}

\newcommand{\nab}{\boldsymbol{\nabla}}

\newcommand\R{\mathbb{R}}

\newcommand\N{\mathbb{N}}

\newcommand{\Exp}{\mathbb{E}}

\newcommand{\mbs}[1]{\ensuremath{\boldsymbol{#1}}}
\newcommand{\bcdot}{\mbs{\cdot}}

\newcommand{\Df}{\mathrm{\mathbb{D}}} 
\newcommand{\D}{\mathrm{\mathbb{D}}^{\tiny \circ}}
\newcommand{\dom}{{\mathcal S}}
\newcommand{\sdbt}{\bsigma_t \dif \W_t} 
\newcommand{\sodbt}{\bsigma_t \circ \dif \W_t} 
\renewcommand*{\div}{\nabla\bcdot} 
\newcommand{\adv}{\!\bcdot\!\nabla} 

\newcommand{\covariation}[2]{\dif_t\biggl\langle \int_0^{\bcdot} #1\,,\int_0^{\bcdot} #2 \biggr\rangle_{\!\!t}}

\begin{document}
\pagestyle{headings}
\setcounter{tocdepth}{4}
\title{Variational principles for fully coupled stochastic fluid dynamics across scales}

\author[1]{Arnaud Debussche}
\author[1]{Etienne M\'emin}
\affil[1]{Univ Rennes, CNRS, Inria, IRMAR UMR 6625, 35042 Rennes, France}

\maketitle

\begin{abstract}
This work investigates variational frameworks for modeling
stochastic dynamics in incompressible fluids, focusing on 
large-scale fluid behavior alongside small-scale stochastic 
processes. The authors aim to develop a coupled system of 
equations that captures both scales, using a variational
principle formulated with Lagrangians defined on the full
flow, and incorporating stochastic transport constraints. The
approach smooths the noise term along time,
leading to stochastic dynamics as a regularization
parameter approaches zero. Initially, fixed noise terms are
considered, resulting in a generalized stochastic Euler
equation, which becomes problematic as the regularization
parameter diminishes. The study then examines connections
with existing stochastic frameworks and proposes a new
variational principle that couples noise dynamics with
large-scale fluid motion. This comprehensive framework
provides a stochastic representation of large-scale
dynamics while accounting for fine-scale components. Our main result is that the
evolution of the small-scale velocity component is governed
by a linearized Euler equation with random coefficients,
influenced by large-scale transport, stretching, and
pressure forcing.
\end{abstract}





\section{Introduction}
Today and in the foreseeable future, no numerical simulation can realistically model the entire array of interacting multiple scales present in fully developed turbulent flows. This is particularly evident in geophysical flow circulation, which spans from the scale of the Sun's heating (approximately 10,000 km) down to the turbulence dissipation scale (approximately 1\,mm). Energy transfers occur towards smaller scales, or conversely, in the opposite direction, influenced by rotation and stratification \cite{Smith-et-al-PRL-96, Smith-Waleffe-POF99, Xia-et-al-11}.  The overlooked sub-grid processes in turbulent flow dynamical systems must be adequately considered to ({\em i}) faithfully represent at coarse resolution (relatively to the dissipation scale) the spread of an ensemble of realizations from a set of imperfect or unknown initial conditions, ({\em ii}) achieve accurate energy transfers, and ({\em iii}) define stable numerical simulations.

The first point is dynamical and potentially impacts severely the  chaotic nature of the system. The second one is physical and impedes the representation of key processes in the energy transfer across scales, while the third is numerical and can lead without caution to numerical instabilities, potentially causing blow-ups, the generation of spurious artifacts or enforcing over-smooth solutions. All three of these effects are crucial in large-scale numerical models for forecasting, data assimilation, and data analysis. Although they may seem quite different in nature, they are challenging to disentangle in practice.

Foremost, the impact of sub-grid scale parameterization on large-scale transport is one of the primary sources of error and uncertainty in simulations of geophysical flows. Sub-grid modeling addresses fundamental issues ranging from the effects of turbulence to the practical design of numerical schemes for computational simulations. This is inevitable since simulating  geophysical flows on large bassin at the Kolmogorov scale is entirely unachievable.

In the case of geophysical flows, intermittent flow-coupled forcing and small-scale processes resulting from thermodynamic effects, species mixing, or biogeochemistry create highly complex systems that are exceedingly difficult to model deterministically. Probabilistic modeling appears to be the most viable approach, especially when priorities include reducing resolution and computational costs, as well as accurately representing uncertainties and their dynamics.

Furthermore, concerning data assimilation and ensemble forecasting, a narrow spread of realizations can pose challenges in coupling data with the system's dynamics. In such cases, observations that deviate significantly from the ensemble cannot effectively correct the ensemble evolution. Given the increasing prominence of ensemble methods and the necessity to simulate multiple realizations covering plausible scenarios, there is a growing need to explore stochastic modeling capable of accurately representing dynamics uncertainties.

Several strategies for introducing randomness into flow dynamics or climate models have been proposed in the literature in recent years \cite{berner2017stochastic,Gottwald2017}. In this work, we first present a specific approach derived from the stochastic transport of fluid parcels \cite{memin2014fluid}. This approach, termed modeling under location uncertainty (LU), has been demonstrated to be versatile and enables the derivation of flow dynamics from classical conservation laws \cite{Bauer2020ocemod, Brecht-et-al-JAMES-22,chandramouli2018,Chandramouli-JCP-20,chapron2018large, Kadri2017,Resseguier-GAFD-II-17,Resseguier-GAFD-III-17}. Furthermore, unlike settings constructed with empirical random forcing, LU incorporates an inherent mechanism to prevent uncontrolled variance growth. This energy conservation property is advantageous in creating a stochastic system that accurately represents large-scale versions of the deterministic dynamical system \cite{Bauer-et-al-JPO-20,chapron2018large}.

From the point of view of Statistical Physics, this balance can be interpreted as an instance of a fluctuation-dissipation relationship between the noise term and stochastic diffusion. Recently, a mathematical analysis of LU Navier-Stokes equations has been performed \cite{Debusshe-Hug-Memin-2023}. Beyond demonstrating the existence of weak (probabilistic) solutions in 3D -- with uniqueness in 2D -- it was shown that this stochastic model tends toward the deterministic equation as the noise vanishes. This provides  physical consistency to this formal setting.

In essence, LU shares similarities with another approach known as stochastic advection by Lie transport (SALT), derived from a variational formulation, as initially proposed in \cite{holm2015variational}; note that originally, LU is derived from a distribution form of Newton's second law. Readers may also refer to \cite{Street-Crisan-23} for extensions of such variational principles driven by martingale.  {We briefly review this stochastic variational setting. It does not seem possible to relate the LU formalism to a variational approach. However, we show that this is possible at the price of some approximations. Both LU and SALT propose fluid equation with a transport type noise. The analysis of such type of noise is the subject of a very intense research work in the literature. The aptitude of such noise to regularize partial differential equations (PDE's) has been explored in different setting, the justification of stochastic representation of flow dynamics models as well as their analysis are currently investigated by several groups (see for instance \cite{Agresti-et-al-2022,Brzezniak-Slavik-2021,Carigi-Luongo-2023,crisan2019solution,Debusshe-Hug-Memin-2023,Debussche-Pappalattera2023,Flandoli-Galeati-Luo-2021,Flandoli-Luo-2021,Flandoli-Pappalettera-2023,Flandoli-Russo2023,Galeati-Luo-2020,Galeati-Luo2023, Goodair-et-al2022,Lang2023} and references therein for recent publications).

Nevertheless, the primary aim of this study is to propose a variational principle that enables the derivation of dynamics for the small-scale components of the flow. In both LU and SALT, the small scales are described by a spatially dependent white noise in time, which is given {\it a priori}. Several strategies have been proposed within the LU framework to incorporate noise explicitly dependent on the solution  \cite{Bauer-et-al-JPO-20,Brecht-et-al-JAMES-22,Resseguier-et-al-2020,Tucciarone2023}, or driven by dynamics extracted from data \cite{Li-et-al-2022}.  However, no direct derivation of dynamics for the small scales has yet been achieved within the various stochastic frameworks proposed in the literature. We will show that such a variational principle can be formulated to derive dynamics for the set of functions on which the noise is decomposed. Our main result should be seen as complementary to the stochastic modeling of large-scale dynamics. Once a stochastic framework --such as LU or SALT-- has been chosen for the large scales, the dynamics of the noise can then be determined according to our new formalism.

This dynamics takes the form of a linearized Euler equation, involving transport and stretching by the large-scale flow, accompanied by a small-scale pressure forcing. It is important to emphasize that this dynamics does not result from a linearization but rather follows from the variational derivation. In this derivation, we will demonstrate that a Stratonovich-like noise is essential for capturing correlations between the small and large scales. However, transitioning to an equivalent It\^{o} form is necessary, as it is better suited for explicit computations.

The next section is devoted to a rigorous description of the white noises used in our stochastic large-scale modeling. We then present the LU formalism and the variational principle leading to the SALT stochastic equations. Regularizing the noise provides greater flexibility in the variational principle, and we end Section \ref{Sec2} with an attempt to recover the LU formalism from a variational principle that differs from SALT in two ways: (i) the kinetic energy includes contributions from both large and small scales, whereas in SALT it only accounts for the large scales, and (ii) the small scales are modeled by a time-correlated noise rather than a white noise. In all these derivations, it is crucial that the stochastic products are correctly interpreted, and we carefully describe the differences between It\^{o} and Stratonovich products.

With all these tools in hand, we present our new variational principle, which allows us to derive the dynamics of the noise characteristics, evolving according to a linearized Euler equation. In this derivation, it is crucial that the regularized noise corresponds to an approximation of a Stratonovich product, ensuring that correlations are properly accounted for. However, It\^{o} products are much more convenient for computations, and we show how to transition from one to the other. We then observe that this approach extends previous classical modeling efforts, which were originally proposed based on purely heuristic arguments.

\section{Stochastic models for large scales}
\label{Sec2}
\subsection{Description of the noise}
Like many large scale flow dynamics representations that describe the flow in terms of a large-scale smooth velocity component and a fluctuation component (with respect to an averaging that must be specified) the models described here start  from a decomposion of the flow in terms of a smooth-in-time Lagrangian velocity component and a highly oscillating zero-mean random component, 
\begin{equation}\label{eq:dX}
\dif \XX_t = \w (\XX_t, t)\, \dif t + \bsigma_t (\XX_t, t)\, \dif \W_t.
\end{equation}
The random fluctuations are modelled on a canonical stochastic basis defined as the quadruple $(\Omega,\mathcal{F},(\mathcal{F}_t)_{t\geq 0},\mathbb{P})$, where $\Omega$ is a set, $\mathcal{F}$ is a sigma algebra, $(\mathcal{F}_t)_{t\geq 0}$ is a right-continuous filtration and $\mathbb{P}$ is a probability measure.  In this decomposition $\XX\,:\, \R^+\times \Omega \to \mathcal {S}$  is the Lagrangian displacement defined within the bounded domain $\mathcal{S} \subset \R^d\ (d = 2\ \text{or}\ 3)$ with smooth boundary, and $\w\,:\, \R^+\times \mathcal{S}\times  \Omega \to \mathcal {S}$  denotes the large-scale velocity that is both spatially and temporally correlated, while $\bsigma\dif \mbs{W}$ is the unresolved component, which is assumed uncorrelated in time and  correlated in space. This latter term is built from a cylindrical Wiener process $\W$ on $H=L^{2}(\mathcal{S} , \R^{d})$, the space of square integrable functions on $\mathcal S$ with values in $\R^d$ \cite{DaPrato} and through a time dependent  integral covariance operator 
$\bsigma_t$  defined for each $\omega \in \Omega$ from a bounded and symmetric positive kernel $\widehat{\bsigma}$:
\[
\bigl(\bsigma_t \boldsymbol{f}\bigr)(x)\;  := \; \int_{\mathcal{S}} \widehat{\bsigma}(x,y,t) \; \mbs{f}(y)\; \mathrm{d}y, \; \mbs{f}\in H.
\]
Since the correlation kernel is bounded in $x$, $y$ and $t$, the operator $\bsigma_t$ maps $H$ into itself and is Hilbert-Schmidt. From the spectral theorem for compact, self-adjoint operator the noise component can be conveniently written as the spectral decomposition -- with explicit dependence on the parameters $(x,t,\omega)$, where $\omega$ denotes randomness:
\[
\bsigma_t \W_t(x,\omega)  = \sum_{i\in \N} \beta^i_t(\omega) \boldsymbol{\varphi}_i(x),
\]
where $(\beta_i)_{i\in\N}$ is a sequence of independent standard Brownian motions on the stochastic basis $(\Omega , \mathcal{F},  (\mathcal{F}_t)_{t\geq 0}, \mathbb{P} )$ and $(\mbs{\varphi}_i)_{i\in\N}$ are the correlation operator eigenfunctions scaled by their eigenvalues, $\lambda_i^{1/2}$. In addition, we assume that the operator-valued process $\{\bsigma_t(\cdot)\}_{0\leq t \leq T}$ is stochastically integrable, that is $\Exp\bigl[ \sum_{j\in\mathbb{N}} \lambda_j (\omega, t) \bigr]< \infty$. As such, under the probability measure $\mathbb P$, the stochastic integral $\bigl\{ \int_0^t \bsigma_s \dif \W_s (\cdot)\bigr\}_{0\leq t\leq T}$ is a $H$-valued process of zero mean,  $\Exp_{\mathbb P} \int_0^t \bsigma_s \dif\W_s(\cdot) =0$, and of bounded variance, $\Exp_{\mathbb P}\bigl [\|\int_0^t \bsigma_s \dif\W_s(\cdot)\|^2\bigr]< \infty$.
We denote by $L^2\bigl(\Omega, L^2(\dom)\bigr)$ the Hilbert space of functions of $L^2(\dom)$  of bounded  variance, equipped with inner product $(f,g)_{L^2(\Omega)} = \Exp (f,g)_{L^2(\dom)}$. 
Divergence-free or divergent noise can be considered \cite{Bauer-et-al-JPO-20} but, to lighten the presentation, in this paper we consider only divergence-free noises. This yields a divergence free constraint on the kernel:
$$
\div \hat \bsigma(x,y,t)=0, \; x,y\in \mathcal  S, \; t\ge 0.
$$
Associated with $\boldsymbol{\sigma}_t$, we define the (matrix)  tensor $\mbs{a}_{ij}(x,t)$ as
\begin{equation}
\mbs{a}_{ij}(x,t)=\int_\Omega\boldsymbol{\check{\sigma}}^{ik}(x,x',t) \boldsymbol{\check{\sigma}}^{kj}(x',x,t) \dif x' \; = \; \displaystyle \sum_{k=0}^{\infty} \boldsymbol{\varphi}^{i}_{k}(x,t) \, \boldsymbol{\varphi}^{j}_{k}(x,t).
\label{def-var}
\end{equation}
This quantity corresponds in the general case to the quadratic variation of the noise 
\begin{equation}
\mbs{a}_{ij} (x,t) \dif  t= \covariation{\bsigma_s\dif \W_s^{i}(x)}{\bsigma_s \dif \W_s^{j}(x)}.
\end{equation}
The quadratic variation (whose definition is briefly recap in \ref{sec:bracket}) is a finite variation process when the correlation operator,  $\boldsymbol{\sigma}_t$, is random. For a deterministic operator, it can be understood  as the 
  {one-point} covariance tensor:
\begin{equation}
    \mbs{a}_{ij}(x,t)\dif t=\Exp\left(\left(\bsigma_t \dif \W_t\right)^{i}(x)\left(\bsigma_t \dif\W_{t}\right)^{j}(x)\right),
    \label{eq:a}
\end{equation}
and for that reason we refer to $\mbs a$ as the variance tensor by abuse of language.

\subsection{Location uncertainty}\label{LU}
\paragraph{The Stochastic Reynolds transport theorem}
The LU  formalism is based on the Stochastic Reynolds transport theorem (SRTT), which generalizes the classical Reynolds transport Theorem to Lagrangian particle evolving according to \eqref{eq:dX}.
It has been introduced in \cite{Mikulevicius04,memin2014fluid,Resseguier-GAFD-I-17}. This theorem provides the rate of change of a random scalar $q$ within a volume $V(t)$,  transported by the stochastic flow \eqref{eq:dX}. For general unresolved flows the SRTT reads
\begin{subequations}
\begin{align}
&\dif\, \Big( \int_{V(t)} \!\!\!\!q (x, t)\, \dif x \Big) = \int_{V(t)} \!\!\! \big( \Df_t q + q \div (\w - \bus)\dif t \big)\, \dif x , \label{eq:SRTT} \\
&\Df_t q = \dif_t q + (\w - \bus) \adv q\, \dif t + \sdbt \adv q - \frac{1}{2} \div (\mbs{a} \nabla q)\, \dif t, \label{eq:D-STO} 
\end{align}
\end{subequations}
where $\dif_t q (x, t) = q (x, t + \dif t)  - q (x, t)$ stands for the forward time-increment of $q$ at a fixed point $x$. The operator $\Df_t$ is introduced as the stochastic transport operator \cite{memin2014fluid,Resseguier-GAFD-I-17}, and plays the role of the material derivative. Recall that $\w$ is the  large-scale velocity used in \eqref{eq:dX} and $\mbs{a}$ is defined in \eqref{def-var}. 

In this expression, the resulting advection is characterized by an effective transport component, modified by a drift which can be writen in the case of a divergence-free noise considered here:
\begin{equation}
\label{def-ISD-I}
\ws = \frac{1}{2}\div \mbs{a},
\end{equation} 
referred to as the It\^{o}-Stokes drift (ISD) in \cite{Bauer-et-al-JPO-20}. It represents the resulting statistical effect of the small-scale inhomogeneity action on the transported quantity. For homogeneous divergence-free noises, this quantity is null, i.e. the variance tensor $\boldsymbol{a}$ is constant in space due to homogeneity. 
The third term in \eqref{eq:D-STO}, is intuitive, and corresponds to the advection of scalar $q$ by the small-scale velocity component. This term continuously backscatters energy into the tracer energy through its quadratic variation.  This gain of energy brought by the noise is then exactly compensated by the loss associated with the last diffusion term. This diffusion term is reminiscent of a generalized Boussinesq eddy viscosity -- i.e the variance noise tensor plays the role of a viscosity (note that it has the unit of a viscosity in $m^2/s$). In the stochastic transport operator expression \eqref{eq:D-STO} the stochastic product has to be interpreted in It\^o sense. For the following, it is insightful to write its form with a Stratonovich product.
\paragraph{Stratonovich expression of the SRTT}
The Stratonovich integral has the great advantage to be associated with classical (deterministic) calculus rule. On the other hand, it does not have the martingale property of the It\^o integral. In particular, it is  not of zero mean. Both stochastic integral are nevertheless equivalent in the sense that one can move from one to the other under the hypothesis of regular enough stochastic processes for the Stratonovich integral to exist. It\^o integral is defined for very mild conditions (adaptability with respect to a filtration and square integrability) while the Stratonovich integral requires stronger regularity conditions. 
As described in \cite{Bauer-et-al-JPO-20} the stochastic flow \eqref{eq:dX} and the transport operator can be turned in terms of the following equivalent expressions (with some regularity conditions on the transported quantity) 
\begin{equation}\label{eq:Stratonovich-dX}
\dif \XX_t = \w \dif t - \frac{1}{2} \div \boldsymbol{a} \dif t + \bsigma_t \circ \dif \W_t,
\end{equation}
which involves a Stratonovich expression of the stochastic integral, specifically indicated by the $\circ$ symbol. Again, in the Stratonovich case, the noise term is not anymore of zero expectation. Since the noise is supposed to represent the small scales of fluid and since these are expected to have a zero average, another drawback of the Stratonovich form is that the noise does not represent the small scales anymore in  \eqref{eq:Stratonovich-dX}. These are in fact given by the sum of the last two terms in \eqref{eq:Stratonovich-dX}. 
In other words, the average of the Stratonovich noise is equal to the second term in the right hand side of \eqref{eq:Stratonovich-dX} and this is part of the large scales of the fluid. 

With this expression of the flow the transport operator corresponds now to the Lagrangian (material) derivative associated to the effective transport velocity \cite{Bauer-et-al-JPO-20}: 
\begin{equation}
\label{transp-op}
\D_t q = \dif_t\circ q + (\w - \ws) \adv q\, \dif t + \sodbt \adv q,
\end{equation}
where $\dif_t\circ q\defin \theta\bigl(x, t + (dt /2)\bigr)- \theta\bigl(x, t - (dt /2)\bigr) $  
stands for the centered time increment. 

\paragraph{LU Euler equations}

Using the conservation of mass and the second law of Newton with the SRTT, we obtain the LU form of the stochastic Euler equations for the evolution of the large scales of an incompressible fluid in a domain 
$\mathcal S$ in terms of Ito product:
\begin{equation}
	\begin{aligned}
		&\dif_t \bu + (\bu - \bus) \adv \bu\, \dif t + \sdbt \adv \bu - \frac{1}{2} \div (\mbs{a} \nabla \bu)\, \dif t= -\nab \dif p ,\\
		& \div \bu =0.
	\end{aligned}
	\label{Euler_LU}
\end{equation}
It can be rewriten with Stratonovich product:
\begin{equation}
	\begin{aligned}
		& \dif_t\circ \bu + (\bu - \bus) \adv \bu\, \dif t + \sodbt \adv \bu= -\nab \dif p,\\
		& \div \bu =0.
	\end{aligned}
	\label{Euler_strato}
\end{equation}
\subsection{Stochastic Advection by Lie Transport}

Another approach to derive stochastic Euler equations for a large scale velocity is to use a variational principle. An easy and simplified way to describe this approach in the deterministic setting is to introduce the energy functional 
\begin{equation}
	S(\bv,\rho,p,\lambda)=\int_{t_1}^{t_2} \bigl(\ell(\bv,\rho) + \bigl\langle \lambda\,, \partial_t  \rho  + \nab \bcdot (\bv\, \rho) \bigr\rangle_{L^2(\dom)}\bigr)\dif t.
\end{equation}
Here $\lambda$ is a scalar Lagrange multiplier that enforces the density $\rho$ to be transported. By duality a constraint on the transport by the flow of any scalar (associated with a density multiplier) could be added.  The angle brackets $\langle\,\cdot\,,\,\cdot\,\rangle_{L^2(\dom)}$ denote the $L^2$ pairing over the domain $\mathcal {S}$ and $t_1$ and $t_2$ denote two arbitrary times. In this study, we explore a simple fluid model only.

The first part of the functional is, in the case of the Euler equations, the kinetic energy:
 $$
    \ell(\bv,\rho) = \int_{\mathcal {S}}  \bigl( \frac{1}{2} |\bv|^2\rho\bigr)\,\dif \mathcal {S}.
 $$
 The second part is there to encode the conservation of mass. The critical points of $S$ satisfy the compressible Euler equations. One recovers the incompressibility conditions thanks to the introduction of a supplementary constraint:
\begin{equation}
	S(\bv,\rho,p,\lambda)=\int_{t_1}^{t_2} \bigl(\ell(\bv,\rho)  - \bigl\langle p, \rho-1\bigr\rangle_{L^2(\dom)} + \bigl\langle \lambda\,, \partial_t  \rho  + \nab \bcdot (\bv\, \rho) \bigr\rangle_{L^2(\dom)}\bigr)\dif t.
\end{equation}
Thus, constraints both on density and scalar transport are considered. They are usually referred to  as Clebsh constraints. Importantly, let us note that in order to guaranty a velocity solution that includes both a gradient field and a solenoidal component -- consistent with the Helmholtz decomposition--  a $d$-dimensional  transport constraint  on a vector-valued passive quantity, $\mbs \theta$,   must to be added to the Clebsh constraints \cite{Cotter-Holm-09, Street-Crisan-23}. This constraint takes the form
\[
\bigl\langle \mbs m\,, \partial_t  \mbs \theta  + \bv \bcdot\nab  \mbs \theta) \bigr\rangle_{L^2(\dom)}, 
\]
where  $\mbs m$ is Lagrange multiplier. Neither $\mbs \theta$ nor $\mbs m$ appears in the Lagrangian. Since this constraint ultimately does not alter the resulting equations of motion, it will be omitted in the following for the sake of simplicity.

In the stochastic case, we consider, as in the LU framework, a velocity decomposed into a large scale smooth part and a small scale part represented by a white noise term \footnote{Note in the expression below, it is implicit that one is working in an Eulerian framework, contrary to \eqref{eq:dX} which is a Lagrangian expression. It follows that the stochastic product below can be It\^o or Stratonovich, there is no difference here.}:
\begin{equation}\label{e13}
	\dif \bv= \bu\dif t +\sdbt
\end{equation}
with the same notations as in section \ref{LU}. An immediate difficulty is that the kinetic energy is not well defined since a white noise is not a function. One of the main point in the SALT modelling is to replace the kinetic energy of the whole flow by the kinetic energy of the large scales only. Then the other terms can be defined in natural way provided $u$, $\rho$, $\lambda$, $p$ are nice stochastic processes (semimartingales) and once it has been decided which sense should be given to the product 
$\bv\,\rho$. The second main point of SALT is to interpret this product as a Stratonovich one. 
In other words,  one considers the functional:
\begin{multline}
	S(\bu,\rho,p,\lambda)=\int_{t_1}^{t_2} \bigl(\ell(\bu,\rho)  - \bigl\langle p, \rho-1\bigr\rangle_{L^2(\dom)} + \bigl\langle \lambda\,, \partial_t  \rho  + \nab \bcdot (\bu \rho)  \bigr\rangle_{L^2(\dom)}\bigr)\dif t + \\\int_{t_1}^{t_2}\bigl\langle \lambda\,,  \nab \bcdot ( \rho \sodbt)  \bigr\rangle_{L^2(\dom)},
\end{multline}
and, looking for critical points as in section \ref{Sec3} below, one finds the {\it SALT Euler equations}:
\begin{equation}\label{SALT}
	\begin{aligned}
		& \dif_t\circ \bu + \bigl(\bu \bcdot\nab\bigr)\bu\dif t  +\bigl(\sodbt \bcdot\nab\bigr)\bu+ \nab\bigl(\sodbt\bigr)\bcdot \bu
		= -\nab \dif q,\\
		&\nab\bcdot\bu =0,
	\end{aligned}
\end{equation}
where the notation $\nab \bu \bcdot \bv $ stands for $\sum_\ell (\partial_{x_k} u^\ell) v^\ell$. 

We see that these are related to the LU equation with two differences. The last term of the left hand of \eqref{SALT} is not present in LU. It is fundamental for SALT to preserve important quantities such as the heliticity but prevents from energy conservation, which holds for LU. On another hand, SALT formalism does not contain a It\^o-Stokes drift. 

\subsection{Variational principle with correlated noise and full kinetic energy}
\label{Sec3}
LU is derived from the Reynolds transport theorem and Newton's second law. It does not initially ensue from a variational principle, unlike other settings such as SALT or others \cite{Chen-et-al-2022, holm2015variational,  Street-Crisan-23}. All these settings propose slightly different stochastic parameterizations to account for rapidly evolving unresolved-scale effects. Due to their respective differences, these two schemes exhibit different conservation properties: SALT preserves helicity \cite{holm2015variational}, while LU conserves energy \cite{Bauer-et-al-JPO-20, Resseguier-GAFD-I-17}. Interestingly, neither of these two schemes carries the whole set of Euler equation invariants. In that sense, they likely both correspond to approximations of what should be an ideal stochastic representation of the Euler equations.

As an attempt to find a link between variational principles and LU, one may try to take the noise into account in the kinetic energy. Recall that SALT considers a kinetic energy containing only the large scales. Due to the irregularity of the white noise it is not possible to give a meaning to the kinetic energy of a flow with velocity given by \eqref{e13}, a way to give a meaning to the full kinetic energy is to consider smoothed, or correlated, noises. Let us introduce the regularized flow velocity:
\begin{multline}
\label{W-v}
\bv(x,t) =  \bu^* (x,t) +{\bu^{\epsilon}}(x,t), \\\text{ with } \bu^* = \bu -\bus \text { and }\; {\bu^{\epsilon}}(x,t)= \int^{t+\epsilon}_{t-\epsilon}\!\!\!h_\epsilon(t-s) \boldsymbol \xi_i (x,t) \dif \beta^i_s.
\end{multline}
Another advantage of considering such velocity is that a white noise corresponds to an infinite separation of scales between the large and small scales. The above expression represents a large, but not infinite, separation of scales. This is probably more realistic. 

The noise term $\bu^{\epsilon}$ is defined through a regularization with regular functions $h_\epsilon$  of compact support that integrates to one. Function  $h_\epsilon(t) =  1/\epsilon\, h(t/\epsilon)$ is positive on the support $[-\epsilon, +\epsilon]$ and null otherwise. The  functions $\boldsymbol \xi_i $ in \eqref{W-v} are  velocity basis functions such that $\Exp(| \bu^{\epsilon}(t)|_{L^2(\mathcal{S})}^2)<\infty$. Since the regularization function $h$ considers both past and future of the noise, in the limit of a zero time-scale correlation parameter this noise leads to a Stratonovich transport expression.  This known fact will be further justified later on in  section \ref{Section-FVF-DA} and \ref{sec:I-S}. 

Having in mind the discussion in section \ref{LU} concerning It\^o and Stratonovich noise, we know that the large scales of fluid flow are given by $\bu(x,t)=\bu^* (x,t)+ \bus(x,t)$ and not by $\bu^*(x,t)$. Recall that $\bus$ is the It\^o-Stokes drift which depends only on the noise. We therefore write the decomposition of the velocity as:
\begin{equation}
	\label{W-vbis}
	\bv(x,t)  = \bu (x,t)-\bus (x,t) + {\bu^{\epsilon}}(x,t).
\end{equation}

We thus consider the following energy functional $S(\bu,\rho,\lambda)$ 
\begin{equation}
\label{S1}
S(\bu,\rho,\lambda)=\int_{t_1}^{t_2} \!\!\bigl(\ell(\bv,\rho) + \bigl\langle \lambda\,, \partial_t  \rho  + \nab \bcdot (\bv \rho) \bigr\rangle_{L^2(\dom)}\bigr)\dif t, \text{ with } \bv= \bu-\bus + \bu^\epsilon.
\end{equation}
As discussed earlier, this energy should be complemented by an independent vectorial transport constraint, which is dropped here for simplicity reason as it does not change the equation of motion.  



 We are now ready to consider the principle of least action and consider critical points for the energy functional \eqref{S1}. The differential $\delta S(\bu,\rho,\lambda)$ reads
\begin{multline}
\delta S(\bu,\rho,\lambda)= \int_{t_1}^{t_2}\! \bigl(\bigl\langle \frac{\delta\ell}{\delta \bu}-\rho\nab\lambda \,,\delta\bu \dif t\bigr\rangle_{L^2(\dom)} +\left\langle  \partial_t \rho + \nab\!\bcdot\!\bigl( \bv \rho\bigr)\,,  \delta\lambda \dif t \right\rangle_{L^2(\dom)}   \\
+ \bigl\langle\frac{\delta \ell}{\delta \rho}\,,\delta \rho \dif t\bigr\rangle_{L^2(\dom)}
- \left\langle \partial_t \lambda + \bigl(\bv \bcdot\nab\bigr) \lambda\,, \delta \rho\dif t\right\rangle_{L^2(\dom)}\bigr).
\end{multline}
The variations $\delta\bu, \, \delta \rho, \, \delta \lambda$ in the above expression  are arbitrary and vanish at the times $t_1$ and $t_2$ and we have explicitly used the fact that the stochastic integrals are regularized and obey the classical product rule. We have also used the fact that the variations vanish at the endpoints in time, so that  no initial or terminal terms appear. 
Therefore,considering arbitrary perturbation, $\delta\bu$, $\delta \rho$ and $\delta \lambda$, in order to satisfy the variational principle, the following equations must hold almost surely:
\begin{subequations}
\label{S1-F-der}
\begin{align}
\delta \bu: &\qquad \frac{\delta \ell}{\delta \bu}  -\rho\nab \lambda= 0,\label{eq-delta-u}\\
\delta \rho: &\qquad\frac{\delta \ell}{\delta \rho} - \partial_t\lambda - \bv \bcdot\nab \lambda = 0,\label{eq-delta-rho}\\
\delta \lambda: &\qquad \partial_t \rho + \nab \bcdot\bigl( \bv \rho \bigr) = 0 \label{eq-delta-lambda}.
\end{align}
\end{subequations}
 The next step is to eliminate the Lagrange multiplier $\lambda$. To do so, we  compute
\begin{equation}
\begin{aligned}
\partial_t(\rho\nab \lambda) &= (\partial_t \rho)\nab \lambda + \rho\nab\partial_t  \lambda \\
&= -\nab\bcdot\bigl(\bv \rho\bigr)\nab\lambda + \rho\nab\bigl(\frac{\delta \ell}{\delta \rho}\bigr) - \rho\nab(\bv \bcdot \nab \lambda),
\end{aligned}
\end{equation}
 and with equation \eqref{eq-delta-u} we have finally: 
\begin{equation}
\partial_t \bigl(  \frac{\delta \ell}{\delta \bu}\bigr) + (\bv \bcdot\nab) \bigl(  \frac{\delta \ell}{\delta \bu} \bigr)  =  \rho\nab\bigl(\frac{\delta \ell}{\delta \rho}\bigr) - \partial_{x_i} \bv^j  \bigl(  \frac{\delta \ell}{\delta \bu} \bigr)^j  - \div \bv \bigl(  \frac{\delta \ell}{\delta \bu} \bigr).
\label{eq-SEL}
\end{equation}
This Euler-Lagrange equation  provides a general stochastic expression for the large-scale velocity component evolution. Let us now take the concrete exemple of the Euler equation to infer precisely a system of associated stochastic partial differential equations (SPDE).
\paragraph{Euler equations}
The Lagrangian for the Euler equations in a $3$-dimensional domain $\mathcal {S}$ with the Euclidean metric and Cartesian coordinates are obtained with the $L^2$-kinetic energy
\begin{equation}
\label{PA-Euler}
    \ell(\bv,\rho) = \int_{\mathcal {S}}  \bigl( \frac{1}{2} |{\bv}|^2\rho\bigr)\,\dif \mathcal {S}.
\end{equation}
To enforce incompressibility, as in the deterministic case, we include a constraint with a pressure Lagrange multiplier, $p$, that sets the density $\rho$ to a constant. The constrained action $S(\bv,\rho,p,\lambda)$  that we want to minimize is given by
\begin{equation}
S(\bv,\rho,p,\lambda)=\int_{t_1}^{t_2} \bigl(\ell(\bv,\rho)  - \bigl\langle p, \rho-1\bigr\rangle_{L^2(\dom)} + \bigl\langle \lambda\,, \partial_t  \rho  + \nab \bcdot (\bv\, \rho) \bigr\rangle_{L^2(\dom)}\bigr)\dif t.
\end{equation}
The variational derivatives of the Lagrangian are
\begin{equation}
        \frac{\delta\ell}{\delta\bu}= \rho\bv, \text{ and } 
        \frac{\delta\ell}{\delta \rho} = \frac{1}{2}   |\bv|^2.
\end{equation}
The variational derivatives of the Lagrangian with respect to the velocity variable defines the momentum. The variational derivative with respect to the density gives the Bernoulli function. Inserting these expressions into the stochastic Euler-Lagrange equation \eqref{eq-SEL} complemented with the pressure constraint provides the momentum equation. We obtain formally the system of coupled SPDE's 
\begin{equation}
\begin{aligned}
    \partial_t (\rho\bv) + \bigl(\bv \bcdot\nab\bigr)(\rho\bv) + \nab\bv\bcdot (\rho\bv) 
    + \bigl(\nab\bcdot\bv\bigr)\rho\bv &= \rho\nab\left(\frac{1}{2} |\bv|^2\right) - \nab p,\\
    \partial_t\rho + \nab\bcdot\bigl(\bv\rho\bigr) &= 0,\\
    \rho &= 1.
\end{aligned}
\end{equation}
The incompressibility condition implies that the continuity equation reduces to
\begin{equation}
\label{Cont-1}
    \nab\bcdot\bv = 0,
\end{equation}
while the momentum equation reads
\begin{equation}
\label{SME-Euler}
 \partial_t  \bv + \bigl(\bv \bcdot\nab\bigr)\bv     = -\nab p.
\end{equation}

The resulting stochastic system has exactly the same form as the classical Euler equations and shares formally their properties, including the preservation of both energy and helicity. Mathematically, it becomes ill-defined at the zero limit of the regularization parameter $\epsilon$, making numerical handling challenging for small values of this parameter. Furthermore, the system lacks clear scale separation and does not correspond to neither the LU nor SALT solutions.

However, it can be observed that for a constant large-scale velocity, the Euler equation boils down to the balance   
\begin{equation}
 \partial_t \bu^{\epsilon}  - \partial_t \bus +  \bigl((\bu - \bus + \bu^{\epsilon}) \bcdot \nab\bigr) (\bu^{\epsilon} - \bus)= -\nab p^{\epsilon,\mbs a},
 \label{Noise-balance}
\end{equation}
which consists in a Euler type equation on the noise term and on the It\^{o}-Stokes drift with $\bv$ as the velocity transport.
Assuming this balance  still holds for any large-scale velocity component, $\bu$, we obtain the following system
\begin{subequations}
\begin{align}
&\partial_t \bu + \bigl(\bv \bcdot\nab\bigr)\bu = -\nab (p - p^{\epsilon,\mbs a}),\\
 &\partial_t ( \bu^{\epsilon}-\bus)\! + \!
 \bigl((\bu -  \bus + {\bu^{\epsilon}}) \bcdot \nab\bigr) ( \bu^{\epsilon} - \bus)= -\nab p^{\epsilon,\mbs a},
\label{Small-scale-dyn-a}\\
  & \div \bu - \frac{1}{2}\div  \div \mbs a + \div{\bu^{\epsilon}} =0.
  \label{Small-scale-dyn-b}
 \end{align}
 \label{Small-scale-dyn-sep}
\end{subequations}
The extension of the balance  \eqref{Noise-balance} is justifiable assuming there is a time scale separation where the large scale velocity is much slower and smoother than the fine scales. Besides, we can notice that although these equations are formally perfectly defined for regularized noise,  equation \eqref{Small-scale-dyn-a} is badly defined at the limit of $\epsilon\to 0$ because of the terms $\partial_t  \bu^{\epsilon}$ and $( \bu^{\epsilon}  \bcdot \nab \bu^{\epsilon})$ that correspond to ``Brownian acceleration'' terms. 

Considering all the elements gathered so far, the system can now be finally written in terms of the evolution of the large-scale component with incompressibility constraints:
\begin{equation}
    \begin{aligned}
    	&\partial_t \bu + \bigl(\bv \bcdot\nab\bigr)\bu = -\nab (p - p^{\epsilon,\mbs a}),\\
      & \div \bu -\frac{1}{2} \div \div \mbs a + \div{\bu^{\epsilon}} =0.
    \end{aligned}
    \label{S-Euler}
\end{equation}

When $\epsilon$ goes to zero, $h$ tends to the Dirac evaluation function and we get at the limit the following system:
\begin{equation}
\label{SME-Euler-1}
    \begin{aligned}
    	&\dif_t \bu + \bigl((\bu^* \dif t +  \mbs \xi_i \circ \dif \beta^i)\bcdot\nab\bigr)\bu = -\nab (\dif p - \dif p_\sigma),\\
      & \div \bu -\div \bus= \div \mbs \xi_i =0,
    \end{aligned}
\end{equation}
where the stochastic integral should be understood in a Stratonovich sense. Also
$\dif p_\sigma$ is a semi-martingale pressure term.
The above system corresponds exactly to the LU representation of the Euler system. 
As already mentionned, through It\^{o} integration by part formulae it can be immediately checked that this system preserves energy in the same way as the deterministic Euler equation (for adequate boundary conditions) but looses helicity conservation \cite{Bauer-et-al-JPO-20,Resseguier-GAFD-I-17}. Hence, LU corresponds to an approximation of the initial regularized stochastic Euler equations. It cannot be directly obtained from the considered variational principle.

\paragraph{Remark:  SALT with correlated noise}
As already seen, the SALT Euler equations are obtained from a different variational principle in which the following action functional $S(\bu,\rho,p,\lambda)$ is considered:
\begin{equation}
 S(\bu,\rho,p,\lambda)=\int_{t_1}^{t_2} \bigl( \frac{1}{2}\|\bu\|^2  - \bigl\langle p, \rho-1\bigr\rangle_{L^2(\dom)} + \bigl\langle \lambda\,, \partial_t \rho  + \nab \bcdot (\bv\, \rho) \bigr\rangle_{L^2(\dom)} \bigr) \dif t .
\end{equation}
The kinetic energy corresponds to the norm of the large-scale velocity only, while the density is transported by the entire random flow and the product in the transport constraints are expressed with a Stratonovich noise. If one considers a full velocity with a correlated noise as in \eqref{W-vbis} (and no It\^o-Stokes drift), the associated variational derivatives of the Lagrangian are
    \begin{equation}
        \frac{\delta\ell}{\delta\bu} = \rho\bu,\text{ and } 
        \frac{\delta\ell}{\delta \rho} = \frac{1}{2}   |\bv|^2.
\end{equation}
Injecting these variational derivatives in \eqref{eq-SEL} yields the SALT momentum equation with a correlated noise instead of a time white noise: 
\begin{equation}
\begin{aligned}
    &\partial_t \bu + \bigl(\bv \bcdot\nab\bigr)\bu + \nab\bu^{\epsilon}\bcdot \bu
    = -\nab\left(p -\frac{1}{2} |\bu|^2 -\frac{1}{2} |\bu^{\epsilon} |^2\right),\\
    &\nab\bcdot\bv =0,
\end{aligned}
\end{equation}
where the notation $\nab \bu \bcdot \bv $ stands for $\sum_\ell (\partial_{x_k} u^\ell) v^\ell$. 

%
 
 \section{Modelling the evolution of the noise}
 \subsection{Variational formulation for the noise correlation dynamics with a decorrelation assumption}
 \label{Section-FVF-DA}
 
 In the two large scales stochastic models described above, LU and SALT, {\bf the noise term is specified a priori and does not depend on the solution}. In the LU derivation, it is theoretically possible to consider a noise depending on the the current state of the system. But there is no theoretical way to describe  this dependence. Attempts have been made to use a state dependent noise through statistical decomposition approaches, projections  or wavelet basis \cite{Li-et-al-2023, Li-et-al-2022,Tucciarone2023, Tucciarone-25}. However, the variational approach and SALT formalism does not seem compatible with a state dependent noise. 

In this section we adopt a new and different point of view. We aim to construct a variational principle for specifying a proper dynamics for  the noise basis functions $\{\mbs{\xi}_i, i\in \N\}$. 
This new variational principle is built upon the previous variational principle. But instead of trying to find the flow of the large scale as a critical point of the energy functional, our aim is to find the $\xi_i$'s as critical points . Since these do not determine the paths of the flow but only its law, it does not make sense to keep an energy functional which is expressed almost surely and it is very natural to consider the new energy functional as the expectation of the previous one.

From this perspective, we propose a two-stage process. First, the dynamics of the full flow, {\it i.e.} the sum the of the large scales $\bu^*$ and of the  small scales $\bu^\epsilon$ (see \eqref{W-v}), will be assumed to satisfy the pathwise variational principle. These dynamics depend on the noise direction which are then given by our new variational principle. 
%

Once the evolution of the noise is obtained, since the variational principle of section \ref{Sec3} does not give an evolution of the large scales, we have to decide a model for this evolution. We have the choice. Either we make the approximations detailed in section \ref{Sec3} and consider that they evolve according to LU Euler. Or we choose another model such as SALT for instance.
This approach resembles the classical decomposition into large scales and small scales as commonly performed in turbulence modeling.

In the second variational principle expressed in expectation, we need to express all the correlations between the correlated regularized noise considered and the different large-scale components of all the functions involved. To that end, we need to express correlated noise expressions, tending at the limit to Stratonovich representation, to decorrelated, It\^{o} forms of these noise terms. In other words, this comes back to express this second variational principle within a  LU philosophy, where the small-scale components are decorrelated in time and the transport expressed through the It\^{o} form of the stochastic transport operator.
 
 More precisely, for  the noise  functions we consider in the following an action functional, ${\mathbb S}(\rho,\lambda,\mbs{\xi})$ defined as:
 \begin{equation}
\label{Action-Strato-regul}
{\mathbb S}(\rho,\lambda,\mbs{\xi})= \Exp \int_{t_1}^{t_2} \bigl(\ell(\bu,\rho,\mbs{\xi}) + \bigl\langle \lambda \,, \Df_t^\epsilon \rho + \rho \nab \bcdot \bv \bigr\rangle_{L^2(\dom)}\bigr)\dif t,
\end{equation}
 which relies on a regularized stochastic density transport defined as:
\begin{equation}
\Df_t^\epsilon \rho + \rho \nab\bcdot \bv = \partial_t \rho  + \nab \bcdot (\bu \rho) + \\\sum_i \nab\bcdot  \Bigl( \rho(t)\mbs{\xi}_i(t)\int^{t+\epsilon}_{t-\epsilon} h_\epsilon (t-s) \dif \beta^i_s \Bigr)=0.
\label{Trans-Strato-regul}
\end{equation}
 This regularized transport expression involves a correlated noise that will be shown to converge toward a Stratonovich transport expression with Stratonovich noise $\sigma_t \circ \dif \W_t$ (see Appendix B). Let us note that, as above, the large-scale component of the transport, which is assumed to be provided by the first variational principle, includes here directly the It\^{o}-Stokes drift in its expression. It therefore indirectly depends on the noise directions. However, it is known that the evolution of the It\^{o}-Stokes drift is much slower than the noise components.  Therefore, we will 
 neglect this dependence. 
 
 In the following, to fully compute the correlation between the noise and the large-scale velocity component, $\bu$,  we will need to consider an equivalent regularized expression converging toward a transport in  It\^{o} form and involving hence an It\^{o} regularized noise with increments  decorrelated from $\bu(t)$: 
 \begin{equation}
 \sigma_t \W_t^{\epsilon} = \sum_i \int_t^{t+\epsilon} {\tilde h}_\epsilon(t-s)\mbs{\xi}_i(t)\dif \beta^i_s,
  \label{Ito-regul-noise}
 \end{equation}
where $\int_t^{t+\epsilon} {\tilde h}_\epsilon(s)\dif s=1$.

In full generality the noise correlation functions, $\{\mbs{\xi}_i, i \in \N\}$ will be assumed to be stochastic processes of the form
\begin{equation} 
\label{Ito-xi}
   \dif \mbs{\xi}_i = \mbs{\mu}_i \dif t + \sum_j \int_t^{t+\epsilon} {\tilde h}_\epsilon(t-s)\mbs\Lambda_j^i(t) \dif \beta^j_s.
\end{equation} 
Notice that in the noise terms of \eqref{Trans-Strato-regul} and \eqref{Ito-regul-noise} considered above,  the noise correlation functions, $\mbs{\xi}_i$ and $\mbs\Lambda_j^i$, are for simplicity frozen and  evaluated at $t$. In particular, they can come outside of the convolution integral. It can be seen that this introduces a negligible error. 

It can be remarked that  the stochastic density transport in Stratonovich form 
\begin{equation} 
\D_t \rho +\rho\div \bv = \partial_t \rho + \div \bigl( \rho(\bu+  \sigma_t \circ \W^{\epsilon}_t) \bigl)=0,
\end{equation}  
reads in It\^{o} form as 
\begin{equation}
 \partial_t \rho + \div \bigl(\rho (\bu   + \frac{1}{2} \sum_i \mbs\Lambda_i^i )  +  \rho \sigma_t  \W^{\epsilon}_t\bigr) \\- \frac{1}{2} \nab\bcdot \bigl( \mbs{\xi}_i  \div(\mbs{\xi}_i \rho)\bigr) =0.
\end{equation}
Two different possibilities can then be considered. It is possible either to work with the regularized form of the It\^{o} transport formulation (where $\int_t^{t+\epsilon} {\tilde h}_\epsilon(s)\dif s=1$):
\begin{multline}
\label{Trans-Ito-regul}
\partial_t \rho + \div(\bu \rho)  + \frac{1}{2} \sum_i \div(\mbs\Lambda_i^i \rho)   - \frac{1}{2} \nab\bcdot\bigl( \mbs{\xi}_i \div(\mbs{\xi}_i \rho)\bigr) + \\\sum_i  \div \bigl(\rho \mbs{\xi}_i(t) \bigr) \int_{t}^{t+\epsilon} \! {\tilde h}_\epsilon(t-s) \dif \beta_s^i,
\end{multline}
or with a regularized Stratonovich transport, \eqref{Trans-Strato-regul}, (with $\int_{t-\epsilon}^{t+\epsilon} h_\epsilon(s)\dif s=1$) transformed in a It\^{o} form with a decorrelated It\^{o} noise. As shown in Appendix-\ref{sec:I-S} both options are equivalent. 


All the variables, $g$, will be decomposed in terms of a smooth component and a decorrelated noise component, expressed as the sum of correlation functions $ {g_i}$:
 \begin{equation}
g(x,t) =  \bar{g} (x,t) + \!\!\int_t^{t+\epsilon}\!\!\!\!\!\!\!\!\!  {\tilde h}_\epsilon(t-s) {g_i} (x,t) \dif \beta^i_s=  \bar g (x,t) + {g_t^{\epsilon}}(x),
\end{equation}
where the functions $\bar g$ and $g_i$ are  of finite variation.

Relying now on the regularized approximation of the transport expression in its It\^{o} form (with a decorrelated regularized noise, \eqref{Ito-regul-noise}), the action, \eqref{Action-Strato-regul}, considered for the noise correlation functions, is more precisely rewritten for all $j\in \N$ as ${\mathbb S}(\bar \rho, \rho_j, \bar \lambda,\lambda_j ,\mbs{\xi}_j) =$
\begin{equation}
\label{Action-noise-regul-law}
 \Exp\!\! \int_{t_1}^{t_2}\!\!\! \bigr(\ell(\bu,\rho,\mbs{\xi})+ \bigl\langle \lambda \,, \Df_t^\epsilon \rho + \rho \nab \bcdot \bv \bigr\rangle_{L^2(\dom)}\bigr)\dif t.
\end{equation}
The Clebsch constraint related to the density transport can be split and rewritten as :
\begin{multline*}
\Exp\Bigl\langle \bar\lambda\,, \partial_t \bar\rho + \nab\bcdot\bigl(\bu\bar\rho +  \Upsilon_\epsilon \sum_j \mbs{\xi}_j \rho_j\bigr) + \\
\frac{1}{2} \sum_{i}\div ( \mbs \Lambda_i^i   \bar\rho)   - \frac{1}{2}  \sum_{i} \nab\bcdot\bigl( \mbs{\xi}_i \div({\mbs{\xi}_i} \bar \rho)\bigr)  \Bigr\rangle_{L^2(\dom)} + 
\end{multline*}
\begin{multline}
\label{Clebsch-regul-transport}
\sum_j \Exp \Bigl\langle \Upsilon_\epsilon \lambda_j \,, \partial_t \rho_j + \nab\bcdot\bigl(\bu\rho_j + \mbs{\xi}_j\bar \rho \bigr) + \\
\frac{1}{2} \sum_{\ell} \div(\mbs \Lambda_\ell^\ell  \rho_j)   - \frac{1}{2}  \sum_{\ell} \nab\bcdot\bigl( \mbs{\xi}_\ell {\div(\mbs{\xi}_\ell} \rho_j)\bigr) \Bigr\rangle_{L^2(\dom)},
\end{multline}
where $\Upsilon^\epsilon\!\! = \!\int_t^{t+\epsilon} {\tilde h}_\epsilon^2(t-s) \dif s$.

In the above equation, the noise decorrelation and the It\^{o} isometry have been used. We also recall that third order moment of Gaussian random variables cancels. It should be noticed that the temporal derivative term on the third line has been simplified as
\begin{align*}
&\Exp \Bigl\langle  \lambda_j \int_t^{t+\epsilon}\!\!\! {\tilde h}_\epsilon(t-s)\dif \beta^j_s\,,\, \rho_j \partial_t \Bigl(\int_t^{t+\epsilon}\!\!\! {\tilde h}_\epsilon(t-s) \dif \beta^j_s\Bigr) \Bigr\rangle_{L^2(\dom)} \\& \qquad\qquad =\, \Exp \Bigl\langle \lambda_j \rho_j \,,\, \frac{1}{2}\partial_t \int_t^{t+\epsilon}\!\!\! ({\tilde h}_\epsilon(t-s))^2 \dif t\Bigr\rangle_{L^2(\dom)}\\
&\qquad\qquad=\, \Exp \Bigl\langle \lambda_j \rho_j \,,\,  \frac{1}{2}\partial_t\Upsilon_\epsilon \Bigr\rangle_{L^2(\dom)}=0.
\end{align*}

The integration constant reads:  $$\Upsilon^\epsilon\!\! = \!\int_t^{t+\epsilon} {\tilde h}_\epsilon^2(t-s) \dif s\!\sim\! C\epsilon^{-1}\!\gg\! 1.$$

Applying the least action principle and cancelling  out the variational derivatives with respect to the different variables, we obtain the following constraints  (without summation over repeated indices):

\begin{multline}
\Exp\Bigl\langle \frac{\delta \ell}{\delta \bar \rho} - \partial_t \bar \lambda - \bu \bcdot\nab \bar \lambda -\Upsilon_\epsilon\sum_j \mbs{\xi}_j \bcdot \nab \lambda_j \\
- \frac{1}{2} \sum_{i} \nab \bar\lambda \bcdot \mbs \Lambda_i^i  -\frac{1}{2} \sum_{j}  \mbs{\xi}_j \bcdot \nab\bigl ({\mbs{\xi}_j}
\bcdot \nab \bar \lambda\bigr)  \,,  \delta \bar{\rho}\Bigr\rangle_{L^2(\dom)}= 0,\label{eq-LAP-rho-bar}
\end{multline}
\begin{multline}
\Exp\Bigl\langle  \frac{\delta \ell}{\delta \rho_j} -  \Upsilon_\epsilon \partial_t \lambda_j -  \Upsilon_\epsilon \bu\!\bcdot\!\nab\lambda_j - \Upsilon_\epsilon\,  \mbs{\xi}_j\!  \bcdot \!\nab \bar \lambda  -\Upsilon_\epsilon \frac{1}{2} \sum_{\ell} \nab\lambda_j \!\bcdot\! \mbs \Lambda_\ell^\ell \\
- \Upsilon_\epsilon  \frac{1}{2}  \sum_{\ell}  \mbs{\xi}_\ell\! \bcdot \! \nab \bigl( {\mbs{\xi}_\ell}\bcdot \nab \lambda_j\bigr) \,,  \delta \rho_j\Bigr\rangle_{L^2(\dom)}  = 0,\label{eq-LAP-rho-j}
\end{multline}
\begin{multline}
\Exp\Bigl\langle   \partial_t \bar\rho + \nab\bcdot\bigl(\bu\bar\rho + \Upsilon_\epsilon \sum_j \mbs{\xi}_j \rho_j\bigr) + \frac{1}{2} \sum_{j}   \div(\mbs\Lambda_j^j \bar\rho) \\
 - \frac{1}{2}  \sum_{j} \nab\bcdot\bigl( \mbs{\xi}_j \div ({\mbs{\xi}_j} \nab\bar \rho)\bigr) 
\,, \delta \bar{\lambda} \Bigr\rangle_{L^2(\dom)}= 0,\label{eq-LAP-lambda-bar}
\end{multline}
\begin{multline}
\Exp\Bigl\langle  
\partial_t \rho_j + \nab\bcdot\bigl(\bu\rho_j + \mbs{\xi}_j\bar \rho \bigr) + \frac{1}{2} \sum_{\ell} \div(\mbs\Lambda_\ell^\ell \rho_j)    \\
- \frac{1}{2}  \sum_{\ell} \nab\bcdot\bigl( \mbs{\xi}_\ell \div({\mbs{\xi}_\ell}\rho_j)\bigr)
 \,, \Upsilon_\epsilon \delta \lambda_j \Bigr\rangle_{L^2(\dom)}= 0, \label{eq-LAP-lambda-j}
\end{multline}
\begin{multline}
\Exp\Bigl\langle\! \frac{\delta \ell}{\delta  \mbs{\xi}_j}\! - \!\Upsilon_\epsilon  \bar \rho \,\nab \lambda_j \!  - \!\Upsilon_\epsilon  \rho_j \nab\bar \lambda + \\ 
\underbrace{\frac{1}{2}  \sum_{i}\!  \nab\!\bcdot\! (\bar \rho \mbs{\xi}_i)\nab\bar \lambda - \nab( \nab \bar \lambda \!\bcdot\! \mbs{\xi}_i )\bar \rho}_{A}+ \\
\underbrace{\Upsilon_\epsilon \frac{1}{2}  \sum_{i} \bigl( \nab\!\bcdot\!(\rho_j \mbs{\xi}_i)\nab \lambda_j \!-\! \nab(\nab\lambda_j \!\bcdot\! \mbs{\xi}_i)\rho_j\bigr)}_{B} 
 \,,\delta\mbs{\xi}_j\!\Bigr\rangle_{L^2(\dom)}\!\!\!\!\!\! =0,\label{eq-LAP-xi}
 \end{multline}
 \begin{equation}
\Exp \Bigl\langle \frac{1}{2}\bigl(\bar \rho\, \nab \bar \lambda  +  \Upsilon_\epsilon  \sum_{i} \rho_i \nab\lambda_i  \bigr)\,,\delta \mbs\Lambda^j_j\Bigr\rangle_{L^2(\dom)} =0 \label{eq-Lambda}.
\end{equation}

Taking the time derivative  of \eqref{eq-LAP-xi} we have:
\begin{equation}
\begin{aligned}
\label{eq-der-temp-xi}
 &\Exp \Bigl\langle \Upsilon_\epsilon ^{-1}\partial_t \bigl(\frac{\delta \ell}{\delta  \mbs{\xi}_j} \bigr)-    \partial_t \bar{\rho} \nab\lambda_j  -  \bar{\rho} \nab\partial_t \lambda_j - \partial_t \rho_j \nab\bar \lambda - \rho_j \nab\partial_t \bar \lambda  \\
 &\qquad+\Upsilon^{-1}_\epsilon \frac{1}{2}  \partial_t \sum_{i} \bigl( \div (\bar \rho \mbs{\xi}_i)\nab\bar \lambda- \nab( \nab \bar \lambda \bcdot \mbs{\xi}_i )\bar \rho\bigr)\\
&\qquad\qquad + \frac{1}{2}   \partial_t  \sum_{i} \bigl( \nab\!\bcdot\!(\rho_j \mbs{\xi}_i)\nab \lambda_j - \nab(\nab\lambda_j \!\bcdot\! \mbs{\xi}_i)\rho_j\bigr)\,,\delta\mbs{\xi}_j\!\Bigr\rangle_{L^2(\dom)}\!\!\!= 0.
\end{aligned}
\end{equation}
The expression of the temporal derivatives can now be injected in this expression to get an evolution equation of the noise correlation functions. 
In the next section we infer more precisely the evolution equations associated to specific case of the Euler equations.
\subsection{Euler equations}
As previously, the Euler pathwise action, $S(\bu,\rho,p,\lambda)$, is defined by the $L^2$-kinetic energy of the whole flow, $\ell(\bv,\rho)$, together with  density transport  and incompressibility constraints, \eqref{PA-Euler}.
\begin{equation}
    \ell(\bv,\rho) = \int_{\mathcal {S}}  \bigl( \frac{1}{2} |\mathbf{\bv}|^2\rho\bigr)\,\dif \mathcal {S}.
\end{equation}
As explained above, this pathwise action is used to derive the evolution of the full flow. 
We complement it with the expectation of this action to derive the dynamic of the nosie directions. 
We thus consider the averaged action functional, ${\mathbb S}(\mbs{\xi}_i,\bar \rho, \rho_i, \bar p, p_i, \bar \lambda, \lambda_i)$, defined equation \eqref{Action-noise-regul-law}, along with an incompressibility constraint. 
\begin{equation}
{\mathbb S}(\mbs{\xi}_i,\bar \rho, \rho_i, \bar p, p_i, \bar \lambda, \lambda_i)=  \Exp\!\!\int_{t_1}^{t_2}\!\bigl( \ell(\bv,\rho)  - \bigl\langle p, \rho-1\bigr\rangle_{L^2(\dom)}  + \bigl\langle\Df_t^\epsilon \rho + \rho \nab \bcdot \bv \bigr\rangle_{L^2(\dom)}\bigr)  \dif t.
\end{equation}

Note that, if, as for SALT, we consider the kinetic energy of the large scales only in the above actions, we miss the correlations of the small and large scales and we are not able to derive an evolution for the noise components. 

Let us recall that the variational derivatives of the Lagrangian with respect to the large-scale velocity and density are
\begin{equation}
\label{der-u-E}
        \frac{\delta\ell}{\delta\bu}= \rho\bv, \text{ and }
        \frac{\delta\ell}{\delta \rho} = \frac{1}{2}   |\bv|^2,
\end{equation}
while for the noise correlation functions we have
\begin{equation}
\label{der-l-xi-decor-Euler}
        \Exp\bigl\langle \frac{\delta\ell}{\delta \mbs{\xi}_i}\,,\, \delta\mbs{\xi}_i \bigl\rangle_{L^2(\dom)} =   \Exp \bigl\langle\Upsilon_\epsilon (\bar \rho\,\mbs{\xi}_j + \rho_j \bu) \,,\,\delta\mbs{\xi}_j \bigl\rangle_{L^2(\dom)}.
\end{equation}
The pressure variables  give two additional constraints:
\begin{subequations}
\begin{align}
&\Exp\bigl\langle  \bar \rho -1 \,, \delta \bar p \bigr\rangle_{L^2(\dom)}= 0, \label{eq-g}\\
&\Exp\bigl\langle  \Upsilon_\epsilon\rho_i \,, \delta p_i \bigr\rangle_{L^2(\dom)}= 0 \label{eq-h}.
\end{align}
\label{eq-V-Euler-pressure}
\end{subequations}
With these two constraints, it comes out a constant density $\bar \rho=1$ and null correlation functions, $\rho_j$, for  the noise density. From the constraints on density transport \eqref{eq-LAP-lambda-bar} and \eqref{eq-LAP-lambda-j} stem the incompressibility conditions: $\nab\bcdot \bu=0, \nab\bcdot \mbs{\xi}_i =0$. Note that as a consequence of the divergence-free condition on the noise shape functions, the function $\mbs{\mu}_i$ and $\mbs\Lambda_i^j$ are also divergence-free. 

From the pathwise action and its associated stochastic Euler-Lagrange equation, \eqref{eq-SEL}, we obtain  the system of regularized stochastic Euler equations:
\begin{equation}
\begin{aligned}
    &\partial_t \bv + (\bv \bcdot\nab)\bv  = -\nab\left(p-\frac{1}{2} |\bv|^2\right),\\
    & \nab\bcdot\bv = 0.
\end{aligned}
\end{equation}
A shown previously, this system can be approximated at the limit of small $\epsilon$ by the LU dynamics or by another large scale stochastic model.

As for the noise correlation functions, as $\bar \rho =1$ and $\rho_i =0$, from constraint  \eqref{eq-delta-u} and \eqref{der-u-E} we get $ \bar \rho\, \nab \lambda_j =\bar \rho\mbs{\xi}_j$   and  $\bar \rho\,\nab\bar \lambda= \bar \rho\bu$. From equation (\ref{eq-Lambda}), a non zero solution for $\bu$ imposes $\Lambda_i^j=0$, which means the basis function $\mbs{\xi}_j$ are of finite variation. In equation (\ref{eq-der-temp-xi}) we insert the expression of time derivative inferred from constraints (\ref{eq-LAP-rho-bar}, \ref{eq-LAP-rho-j}, \ref{eq-LAP-lambda-bar}, \ref{eq-LAP-lambda-j}).  We then keep only terms in $\bar \rho$, and eliminate terms in $\rho_i$, $\partial_t \bar \rho$ and $\nab \bar \rho$. We also apply the incompressibility condition on $\bu$ and $\mbs{\xi}_j$. 
Equation (\ref{eq-der-temp-xi}) simplifies as 
\begin{equation}
 \Exp \Bigl\langle \Upsilon_\epsilon ^{-1}\partial_t \bigl(\frac{\delta \ell}{\delta  \mbs{\xi}_i} \bigr) -  \bar{\rho} \,\nab\partial_t \lambda_i - \Upsilon^{-1}_\epsilon \frac{1}{2}  \partial_t \sum_{j} \bigl(\nab( \nab \bar \lambda \bcdot \mbs{\xi}_j )\bar \rho \,,\delta\mbs{\xi}_i\Bigr\rangle_{L^2(\dom)} = 0.
\end{equation}
Including the expressions of $\partial_t \lambda_j$ (\ref{eq-LAP-rho-j}):
\begin{equation}
  \partial_t  \lambda_j=  \frac{1}{\Upsilon_\epsilon} \frac{\delta \ell}{\delta \rho_j} -  \bu \bcdot\! \nab\lambda_j -\mbs{\xi}_j \bcdot\! \nab\bar \lambda -   \frac{1}{2}  \sum_{\ell}  \mbs{\xi}_\ell \bcdot\!  \nab \bigl( {\mbs{\xi}_\ell}\bcdot\! \nab \lambda_j\bigr),
\end{equation}
and as $\Exp \bigl \langle \frac{\delta \ell}{\delta \rho_j}, \delta \mbs{\xi}_j\bigr\rangle_{L^2(\dom)} =\Exp\bigl \langle \Upsilon_\epsilon \bu \bcdot{\mbs{\xi}_j}, \delta {\mbs{\xi}_j} \bigl \rangle_{L^2(\dom)}$, we obtain finally a linear Euler type equation for the noise functions dynamics:
 \begin{equation}
 \label{dyn-xi}
 \partial_t{\mbs{\xi}_i} +  (\bu\bcdot \nab){\mbs{\xi}_i} + ({\mbs{\xi}_i}\bcdot \nab) \bu  =  - \nab p^\epsilon,
 \end{equation}
 where the pressure term reads
 \begin{equation}
 \label{noise-function-pressure}
p^\epsilon=       -\mbs{\xi}_i\bcdot \bu + \frac{1}{2} \sum_{\ell} \div \bigl({\mbs{\xi}_\ell}({\mbs{\xi}_\ell}\bcdot{\mbs{\xi}_i} )\bigr) -  \Upsilon^{-1}_\epsilon \frac{1}{2}   \sum_{j} \partial_t( \bu \bcdot \mbs{\xi}_j ).
\end{equation}
 We note that the last pressure term is very small, and can be neglected.
 
 The noise functions are transported and deformed by the large-scale components and a pressure term. The transport of the small scales by the large scales corresponds to Kraichnan's random sweeping hypothesis \cite{Chen-Kraichnan-1989,Kraichnan-1964}. This assumption, also exploited by Tennekes \cite{Tennekes-1975} for turbulent boundary layer models put forward that small eddies (i.e. much smaller than the main energy containing eddies) are transported by an Eulerian large-scale field without any dynamical deformation \cite{Katul-et-al-1995}. The large-scale flow involved  is not the mean flow (which would correspond to Taylor frozen turbulence hypothesis of purely passive turbulence transport) and the absence of dynamical distortion ensues from missing interactions between the small scales and large-scale component, as well as  interactions between the small-scales modes.  The pressure term consists first of  a forcing term depending on the evolution of the angle between the small-scale modes and the large-scale velocity and second, of a turbulent transport term, which is a  third-order product term of the small-scale modes' noise. It is noticeable that this latter term is dominant for spatially small-scale noise correlation functions.  
 
 The pressure term can be explicitly computed or estimated in the usual way by considering the incompressibility condition on the noise basis functions $\mbs{\xi}_j$. More precisely, we obtain the noise pressure $p_i^\epsilon$ through the elliptic equations:
\begin{equation}
\Delta p_i^\epsilon = \nab\bcdot\bigl((\bu\bcdot \nab){\mbs{\xi}_i} + ({\mbs{\xi}_i}\bcdot \nab) \bu   \bigr).
\end{equation}
 Note that also a further simplified equation can be settled through the Leray projector, ${\cal P}$,  onto the space of divergence-free vectors. We get in that case:
 \begin{equation}
  \partial_t{\mbs{\xi}_i} +  {\cal P} (\bu\bcdot \nab{\mbs{\xi}_i} ) +  {\cal P}( \mbs{\xi}_i  \bcdot \nab \bu ) = 0.
 \end{equation}  
 This equation has the advantage of not requiring any boundary condition for the pressure term, which can often be cumbersome to set.

 To sum up, if one chooses LU for the evolution of the large scales,  we obtain the following large scale Euler equation, with an explicit expression of the small-scale component evolution.  
 \begin{equation}
    \begin{aligned}
    	&\partial_t \bu + \bigl((\bu-\bus+\bu^\epsilon) \bcdot\nab\bigr)\bu = -\nab p,\\
      & \div \bu -\frac{1}{2} \div \div \mbs a =0,\\
      &\bu^\epsilon= \int^{t+\epsilon}_{t-\epsilon}\!\!\!h_\epsilon(t-s) \boldsymbol \xi_i (x,t) \dif \beta^i_s, \text{ with }\\
      &\partial_t{\mbs{\xi}_i} +  (\bu\bcdot \nab){\mbs{\xi}_i} + ({\mbs{\xi}_i}\bcdot \nab) \bu  =  - \nab p^\epsilon,\\
      &\nab\bcdot {\mbs{\xi}_i}=0.
    \end{aligned}
    \label{S-Euler-Sum-up}
\end{equation}
Letting $\epsilon \to 0$ we  then obtain the stochastic LU system:
\begin{equation}
\label{SME-Euler-fin}
    \begin{aligned}
    	&\dif_t \bu + \bigl((\bu^* \dif t +  \mbs \xi_i \circ \dif \beta^i)\bcdot\nab\bigr)\bu = -\nab (\dif p - \dif p_\sigma),\\
      & \div \bu -\div \buss= \div \mbs \xi_i =0,\\
      &\partial_t{\mbs{\xi}_i} +  (\bu\bcdot \nab){\mbs{\xi}_i} + ({\mbs{\xi}_i}\bcdot \nab) \bu  =  - \nab q.
    \end{aligned}
\end{equation}
The limit $\eqref{S-Euler-Sum-up}\to \eqref{SME-Euler-fin}$ can be  justified rigorously (see \cite{debussche-hofmanova2023, Debussche-Pappalattera2023}). Physically, \eqref{S-Euler-Sum-up} may be more interesting than \eqref{SME-Euler-fin} since it does not assume a complete decorrelation between the small and large scales.

\subsection{Ornstein-Uhlenbeck process for the regularized noise}
 One simple solution consists in considering an Ornstein-Uhlenbeck (OU) process for the noise, denoted $\mbs{Z}_t^{\epsilon}$ in the following. In that case the regularized noise is defined through the Ornstein-Uhlenbeck semi-group generator. A regularized  noise with an OU process reads:
 \begin{equation}
 \bu^\epsilon =\sigma_t \mbs{Z}_t^{\epsilon}= \sum_i \mbs{\xi}_i(t){Z}_t^{\epsilon,i}  =   \sum_i \mbs{\xi}_i(t)\bigl( e^{-\frac{1}{\epsilon}t} {Z}_{t-\epsilon}^{\epsilon,i} + \int_t^{t+\epsilon}\!\!\!e^{-\frac{1}{\epsilon}(t-s)}\dif \beta^i_s\bigr).
  \label{OU-Ito-regul-noise}
 \end{equation}
 The exponential functional does not have a compact support but this regularization also fully enters within the set of the regularized noises considered in this work. For this noise, the dynamics of the small-scale velocity is then given by:
 \begin{equation}
  \dif_t \bu^\epsilon = -\Bigl(\sum_i \bigl((\bu\bcdot \nab){\mbs{\xi}_i} + ({\mbs{\xi}_i}\bcdot \nab) \bu  + \nab p^\epsilon\Bigr)\mbs{Z}_t^\epsilon\dif t  - \frac{1}{\epsilon}  \sum_i  \mbs\xi_i {Z}_t^{\epsilon,i} -  \sum_i  \mbs{\xi}_i \dif \beta^i_t.
  \end{equation}
In the model described above, the nonlinear small-scale self-interaction is represented by an Ornstein-Uhlenbeck process, and the dynamics of the small-scale velocity components are governed by a linear SPDE. This approach is similar to the scheme proposed in \cite{Majda-et-al-1999,Majda-et-al-2001}, where nonlinear self-interactions are modeled using a linear stochastic operator in the form of an Ornstein-Uhlenbeck process, composed of a damping term and an additive noise term. Our framework fully justifies this form, thereby validating the assumptions made in \cite{Majda-et-al-1999,Majda-et-al-2001} regarding the representation of nonlinear self-interactions. We remark that if the noise functions are assumed stationary ($\partial_t \mbs{\xi}_i =0$ ), we are left with an Ornstein-Uhlenbeck process for the unresolved components of the form: 
  \begin{equation}
  \dif_t \bu^\epsilon =   -\frac{1}{\epsilon}  \sum_i  \mbs\xi_i {Z}_t^{\epsilon, i} -  \sum_i  \mbs{\xi}_i \dif \beta^i_t= -\frac{1}{\epsilon}\bu^\epsilon -  \sum_i  \mbs{\xi}_i \dif \beta^i_t.
  \end{equation}
Such models have been intensively used in ocean modelling and climate science for representing intermittent small-scale boundary layer processes in ocean-atmosphere interactions \cite{Hasselmann-1976,McWilliams-Huckle-2006,Saravanan-McWiliams-1998}. They have been successful in explaining the ubiquitous red spectrum of sea surface temperature \cite{Frankignoul-Hasselmann-1977,Hasselmann-1976}, El Ni\~{n}o Southern Oscillation variability \cite{Kleeman-2011} as well as thermohaline circulation  variability \cite{Griffies-Tziperman-95} and mean current-fluctuations interactions \cite{Delsole-Farrell-96}. The framework derived here can be seen as a generalization of these models, in which we provide and justify a dynamics for the noise modal functions.



\section{Discussion and conclusion}

This study addresses several research issues in turbulence fluid modeling, particularly the accurate representation of multi-scale interactions, including the influence of small-scale turbulence on large-scale ocean dynamics and vice versa. Traditional models often rely on phenomenological or heuristic approaches, which can lack rigor and fail to capture the complexity of these bidirectional interactions. Here, we provide a rigorous derivation that goes beyond these traditional methods.

One significant contribution of this work is the introduction of a stochastic partial differential equation (SPDE) framework with regularized noise terms for modeling both the large-scale velocity components and the dynamics of the small-scale noise terms. The random sweeping hypothesis, initially proposed by Kraichnan \cite{Kraichnan-1964} and further developed by Tennekes \cite{Tennekes-1975} for the atmospheric boundary layer, assumes a passive transport of small scales by large scales. These models extend Taylor's ``frozen'' turbulence scheme, where turbulence is advected by a mean field, by incorporating randomness in the transport velocity field. 
The coupled variational framework proposed in this study goes further by deriving  a stochastic partial differential equation (SPDE) describing the evolution of the large-scale velocity component in tandem with the dynamics of small-scale noise terms. This latter, consisting of a linear Euler-type equation, justifies the description of small-scale dynamics through large-scale random advection as proposed in Kraichnan's sweeping hypothesis. However, it also incorporates additional deformation terms associated with pressure and stretching. 

Our approach also relates to the MTV model \cite{Majda-et-al-1999, Majda-et-al-2001} for stochastic climate modeling and simpler models where noise modal functions are stationary, aligning then with Hasselmann's 1976 model \cite{Hasselmann-1976}. By proposing a formal variational principle, we provide a systematic method to derive governing equations for both large-scale flows and their stochastic perturbations, potentially leading to more accurate and reliable models.

To our knowledge, this is the very first time a formal principle is proposed to derive a stochastic dynamics with an explicit evolution for the small-scale component. Future works will focus on extending this framework to the primitive equations, fundamental in describing ocean dynamics, and to wave-current interactions. This extension is crucial for accurately modeling coastal dynamics, energy dissipation, and nutrient mixing, which are inherently complex due to small-scale features and nonlinear wave interactions.

Overall, our approach bridges a critical gap in the large-scale modeling of turbulent flows, setting the stage for future research that could lead to more accurate predictions by inherently accounting for uncertainties associated with unresolved small-scale nonlinear phenomena.

In future studies, we will undertake a mathematical analysis of the proposed stochastic coupled system with additional friction terms, transforming it into a coupled Navier-Stokes system. The goal will be to demonstrate the existence of (probabilistic) weak solutions, similar to the approach used for the large-scale Navier-Stokes LU equation \cite{Debusshe-Hug-Memin-2023}.


\paragraph*{Acknowledgements:} 
The authors acknowledge the support of the ERC EU project 856408-STUOD and  benefit from the support of the French government ``Investissements d'Avenir'' program ANR-11-LABX-0020-01.

\appendix
\section{Quadratic (co-)variation}\label{sec:bracket}
In stochastic calculus, quadratic covariation (or cross-variance) of two real-valued processes $X$ and $Y$ play a fundamental role. Quadratic variation is a bounded variation process defined as:
\begin{equation}\label{eq:defbracket}
\langle X, Y \rangle_t = \lim_{n\rightarrow 0} \sum_{i=1}^{p_n} (X_{i}^{n} - X_{i-1}^{n})(Y_{i}^{n} - Y_{i-1}^{n}),
\end{equation}
where $0=t_0^n<t_1^n<\cdots<t_{p_n}^n=t$ is a partition of the interval $[0,t]$ and this limit, if it exists, is defined in the sense of convergence in probability.

Assuming that $X$ and $Y$ are two real-valued continuous semimartingales, defined as $X_t = X_0 + A_t + M_t, Y_t = Y_0 + B_t + N_t$ with $M, N$ martingales and $A, B$ finite variation processes, then their quadratic covariation \eqref{eq:defbracket} exists, and is given by 
\begin{equation}\label{eq:semibracket}
\langle X, Y \rangle_t = \langle M, N \rangle_t.
\end{equation}
In particular, the quadratic variation of a standard Brownian motion $B$ (as a martingale) is given by $\langle B \rangle_t = t$, the quadratic variation of two bounded variation processes $f$, and $g$ (such as deterministic functions) can be shown to be zero ($\langle f,g\rangle_t=0$), as well as the covariation between a martingale and a bounded variation process ($\langle f,M\rangle_t=0$).
\sectionmark{It\^{o} form of  density transport regularized noise}

The quadratic (co-)variations play an important role in the It\^{o} calculus and its generalization of the chain rule. In particular, they are involved in the It\^{o} integration by parts formula:
\begin{equation}\label{eq:IPP}
\dif (XY) = X \dif Y + Y \dif X + \dif \langle X, Y \rangle_t.
\end{equation}
The quadratic variation of the It\^{o} integrals of two adapted processes with respect to martingale, $M$ and $N$, respectively, is provided by the following important  formula:
\begin{equation}
    \biggl\langle \int_0^{\cdot} \Theta_s \dif M_s\,, \int_0^{\cdot} \Theta'_s \dif N_s \biggr\rangle_t = \int_0^t \Theta_s \Theta'_s \dif \langle M, N \rangle_t 
\end{equation}
This property is involved in the It\^{o} isometry, enabling to express the covariance of two It\^{o} integrals:
\begin{equation}\label{eq:isometry}
\Exp \Big[ \big(\int_0^t f \dif M_s\big) \big(\int_0^t g \dif N_s\big) \Big] = \Exp \Big[ \int_0^t fg \,\dif  \langle M, N \rangle_s \Big],
\end{equation}
where $f$ and $g$ are two adapted processes such that $\int_0^t f^2 \dif \langle M,M \rangle_s$ and $\int_0^t g^2 \dif \langle N,N \rangle_s$ are integrable.

\section{It\^{o} form of the density transport with a regularized  correlated noise}\label{sec:I-S}
\sectionmark{It\^{o} form of  density transport regularized noise}
In this appendix we show that the density transport with regularized correlated noise term (i.e. of Stratonovich type)
\begin{equation}
\Df_t^\epsilon \rho + \rho \nab\bcdot \bv = \partial_t \rho  + \nab \bcdot (\bu \rho) + \\\sum_i \nab\bcdot  \Bigl( \rho(t)\mbs{\xi}_i(t)\int^{t+\epsilon}_{t-\epsilon} h_\epsilon (t-s) \dif \beta^i_s \Bigr)=0,
\label{Trans-Strato-regul-Ap}
\end{equation} 
 up to negligible terms, reads as 
\begin{multline}
\label{Trans-Ito-regul}
 \partial_t \rho + \div(\bu \rho)  + \frac{1}{2} \sum_i \div(\mbs\Lambda_i^i \rho)   - \frac{1}{2} \nab\bcdot\bigl( \mbs{\xi}_i \div(\mbs{\xi}_i \rho)\bigr) + \\\sum_i  \div \bigl(\rho \mbs{\xi}_i(t) \bigr) \int_{t}^{t+\epsilon} {\tilde h}_\epsilon(t-s) \dif \beta_s^i,
\end{multline}
for (It\^{o} type) regularized noise with increments  decorrelated from $\bu(t)$:
 \begin{equation}
 \sigma_t \W_t^{\epsilon} = \sum_i \int_t^{t+\epsilon} {\tilde h}_\epsilon(t-s)\mbs{\xi}_i(t)\dif \beta^i_s,
  \label{Ito-regul-noise-Ap}
 \end{equation}
where the functions, $\{\mbs{\xi}_i, i \in \N\}$ are defined in full generality as stochastic processes of the form
\begin{equation} 
\label{Ito-xi}
   \dif \mbs{\xi}_i = \mbs{\mu}_i \dif t + \sum_j \int_t^{t+\epsilon}{\tilde h}_\epsilon(t-s)\mbs\Lambda_j^i(t) \dif \beta^j_s.
\end{equation} 

More precisely, we show below that the following approximation holds
\begin{equation}
\Df_t^\epsilon \rho + \rho \nab\bcdot \bv \approx \widetilde{\Df}_t^\epsilon \rho + \rho \nab\bcdot \bv,
\label{Approx-Strato}
\end{equation} 
where the approximated density transport $\widetilde{\Df}_t^\epsilon \rho + \rho \nab\bcdot \bv$ reads
\begin{multline}
 \dif_t \rho + \div (\bu\rho)\dif t  + \frac{1}{2} \sum_{ij} \nab\!\bcdot(\mbs\Lambda_i^j b_{ij}^\epsilon \rho) \dif t  \\ -  \frac{1}{2}  \sum_{ij} \nab\!\bcdot\!\bigl(\delta_{ij}^\epsilon \mbs{\xi}_i \nab\!\bcdot\!({\mbs{\xi}_j}\rho)\bigr)\dif t  + \sum_i \nab\!\bcdot\!(\mbs{\xi}_i \rho)\! \int_{t}^{t+\epsilon} \!\!\!\!\tilde h_\epsilon (t-s) \dif \beta^i_s,
\end{multline}
with $\delta_{ij}^\epsilon(t) \overset{a.s}{\longrightarrow} \frac{1}{2} \delta_{ij}$ and $b_{ij}^\epsilon \overset{a.s}{\longrightarrow} \frac{1}{2} \delta_{ij}$ and consequently
\begin{equation}
\Df_t^\epsilon\approx  \widetilde{\Df}^\epsilon_t \quad\text{and}\quad \lim_{\epsilon\to 0}  \widetilde{\Df}_t^\epsilon \longrightarrow \Df_t.
\end{equation}

In order to show this, the noise term in \eqref{Trans-Strato-regul-Ap}  is written for each $i$ as:
\begin{multline}
\label{TS-regul-bis}
\int_0^T \Bigl(\div\bigl(\mbs{\xi}_i(t)\rho(t)\bigr)   \int^{t+\epsilon}_{t-\epsilon} h_\epsilon (t-s) \dif \beta^i_s\Bigr) \dif t =\\\underbrace{ \int_0^T \Bigl(\div\bigl( \mbs{\xi}_i(t) \rho(t)\bigr)     \int^{t+\epsilon}_{t} h_\epsilon (t-s) \dif \beta^i_s\Bigr) \dif t}_{A1} + \\
\underbrace{\int_0^T \Bigl( \bigl(\div\bigl(\mbs{\xi}_i(t-\epsilon)\rho(t-\epsilon)\bigr)    \int^{t}_{t-\epsilon} h_\epsilon (t-s) \dif \beta^i_s\Bigr) \dif t}_{A2} + \\
\underbrace{\int_0^T \Bigl( \div\bigl(\mbs{\xi}_i(t) \rho(t) - \mbs{\xi}_i(t-\epsilon ) \rho(t-\epsilon) \bigr)   \int^{t}_{t-\epsilon} h_\epsilon (t-s) \dif \beta^i_s \Bigr)\dif t }_{B}.
\end{multline}
The two first term $A=A_1+A_2$ on the right-hand side can be approximated as: 
\begin{equation}
A\approx \int_0^T \div \bigl(\mbs{\xi}_i(t) \rho (t)\bigr)    \int^{t+\epsilon}_{t} \underbrace{\bigl(h_\epsilon (t-s) + h_\epsilon(t+\epsilon-s)\bigr)}_{\tilde{h}_\epsilon(t-s)} \dif \beta^i_s \dif t,
\end{equation}
which corresponds to a smoothing of the It\^{o} integral, with kernel, $\tilde h_\epsilon$, such that $\int_0^\epsilon \tilde h_\epsilon =1$. The second right-hand side term of \eqref{TS-regul-bis}, is written $B=B_1 + B_2$ with 
\begin{align}
B_1&= \int_0^T \biggl(\div \Bigl(\mbs{\xi}_i(t)  \bigl( \rho(t) - \rho(t-\epsilon)\bigr)\Bigr)\int^t_{t-\epsilon} h_\epsilon (t-s) \dif \beta^{i}_s \biggr)\dif t\\
&\approx - \int_0^T \biggl(\div \Bigl(  \mbs{\xi}_i(t)\int^t_{t-\epsilon} \biggl[\sum_j  \bigl(\int_{s-\epsilon}^{s+\epsilon} h_\epsilon(s-r)\dif\beta^j_r\bigr) \div\bigl( \mbs{\xi}_j(s)\rho(s)\bigr)\biggr]\dif s  \Bigr)\nonumber\\ 
&\quad\quad\quad\quad\int^t_{t-\epsilon}h_\epsilon(t-\tau) \dif\beta^i_\tau\biggr) \dif t\\
&\approx - \sum_j  \int_0^T \biggl( \int^t_{t-\epsilon} \int^{s+\epsilon}_{s-\epsilon} h_\epsilon (s-r) \dif \beta^j_r  \dif s \div\Bigl( \mbs{\xi}_i(t) \div\bigl(\mbs{ \xi}_j(t)\rho(t)\bigr) \Bigr)\nonumber\\ 
&\quad\quad\quad\quad\int^t_{t-\epsilon} h_\epsilon(t-\tau)\dif \beta^i_\tau \biggr)\dif t \\
& \approx - \sum_j  \int_0^T \delta_{ij}^\epsilon(t) \div \Bigl( \mbs{ \xi}_j(t)\div \bigl(\mbs{\xi}_i(t)  \rho(t)\bigr)\Bigr) \dif t =\tilde B_1, 
\end{align}
where 
\begin{equation}
\delta_{ij}^\epsilon(t) = \int^t_{t-\epsilon} \int_{s-\epsilon}^{s+\epsilon} \!\!\!\!\!\!h_\epsilon(s-r) \dif \beta^j_r \dif s\,  \int_{t-\epsilon}^{t} \!\!\! \!\!\!h_\epsilon(t-\tau) \dif \beta^i_\tau \overset{a.s}{\longrightarrow} \frac{1}{2} \delta_{ij},  
\end{equation}
and hence 
\begin{equation}
\tilde B_1 \overset{a.s}{\longrightarrow} - \frac{1}{2} \nab\bcdot \bigl(\mbs{\xi}_i \div({\mbs{\xi}_i} \rho)\bigr).
\end{equation}
The other term $B_2$ is
\begin{align}
\!\!\!\!\!\!\!B_2 &= \int_0^T \Bigl[\nab\!\bcdot\!\Bigl(\bigl(\mbs{\xi}_i(t) - \mbs{\xi}_i(t-\epsilon)\bigr)\rho(t-\epsilon)\Bigr)\!\!\! \int_{t-\epsilon}^t h_\epsilon (t-s) \dif \beta^i_s \Bigr] \dif t\\ 
&\approx \int_0^T\Bigl[ \nab\!\bcdot\! \Bigl(\sum_j \int^t_{t-\epsilon} h_\epsilon (t-s) \mbs\Lambda^i_j (t) \dif \beta^j_s \,\rho(t-\epsilon)\Bigr)\!\!\!\int^t_{t-\epsilon} h_\epsilon (t-s)\dif \beta^i_s \Bigr]\dif t\\
 &\approx \sum_j  \int_0^T  \nab\!\bcdot\!\bigl(\mbs\Lambda_j^i (t) \rho (t)\bigr)\underbrace{\!\!\Bigl( \int^{t+\epsilon}_{t}  \!\!\! (h_\epsilon(t-s)\dif \beta^j_s \Bigr) \!\!\!  \int_{t}^{t+\epsilon}  \!\!\! h_\epsilon (t+\epsilon-s) \dif \beta^i_s}_{b_{ij}^\epsilon(t)} \dif t \\
 &\approx \sum_j  \int_0^T  b_{ij}^\epsilon(t) \div\bigl(\mbs\Lambda_j^i  \rho(t)\bigr) \dif t,  
\end{align}
with 
\begin{equation}
b_{ij}^\epsilon \overset{a.s}{\longrightarrow} \frac{1}{2} \delta_{ij}. 
\end{equation}
The approximation of density transport $\Df_t^\epsilon \rho+\rho\div \bv$ reads thus as: 
\begin{multline}
\dif_t \rho + \div (\bu\rho)\dif t  + \frac{1}{2} \sum_{ij} \div(\mbs\Lambda_i^j b_{ij}^\epsilon \rho) \dif t   -\frac{1}{2}  \sum_{ij} \nab\bcdot\bigl(\delta_{ij}^\epsilon \mbs{\xi}_i \div({\mbs{\xi}_j}\rho)\bigr)\dif t  + \\ \sum_i \div(\mbs{\xi}_i \rho) \int_{t}^{t+\epsilon}\!\!\! \tilde h_\epsilon (t-s) \dif \beta^i_s.
\end{multline}
and we have consequently:
\begin{equation}
\Df_t^\epsilon\approx  \widetilde{\Df}^\epsilon_t \quad\text{and}\quad \lim_{\epsilon\to 0}  \widetilde{\Df}_t^\epsilon \longrightarrow \Df_t.
\end{equation}


\bibliographystyle{plain}
\bibliography{biblio}

\end{document}